\documentclass[]{spie}  

\usepackage{amsmath,amsfonts,amssymb}
\usepackage{graphicx}
\usepackage[export]{adjustbox}
\usepackage[colorlinks=true, allcolors=blue]{hyperref}
\usepackage{float}
\usepackage{placeins}

\title{Early Telescope Throughput Results from the Collimated Beam Projector at the Vera C. Rubin Observatory}

\author[a]{Nathan Amouroux}
\author[b]{Parker Fagrelius}
\author[a]{Thibault Guillemin}
\author[c]{Fritz Mueller}
\author[d]{Jérémy Neveu}
\author[c, e]{Eli Rykoff}
\author[a]{Thierry Souverin}
\author[f]{Christopher W. Stubbs}
\author[f]{Elana Urbach}
\affil[a]{Laboratoire d'Annecy de Physique des Particules (LAPP), 9 chemin de Bellevue, Annecy, France}
\affil[b]{NSF's NOIRLab, 950 N Cherry Ave, Tucson, AZ 85719, USA}
\affil[c]{SLAC National Accelerator Laboratory, 2575 Sand Hill Rd, Menlo Park, United States}
\affil[d]{Laboratoire Irène Joliot-Curie (IJCLab), 15 Rue Georges Clemenceau, 91400 Orsay, France}
\affil[e]{Kavli Institute for Particle Astrophysics \& Cosmology (KIPAC), 452 Lomita Mall, Stanford, United States}
\affil[f]{Department of Physics, Harvard University, 17 Oxford St, 
Cambridge, MA, 02138 United States}

\authorinfo{Further author information: (Send correspondence to Nathan Amouroux or Thibault Guillemin)\\Nathan Amouroux: E-mail: nathanamouroux@gmail.com \\Thibault Guillemin: E-mail: guillemin@lapp.in2p3.fr}

\begin{document} 
\maketitle

\begin{abstract}
The Vera C. Rubin Observatory LSST requires precise photometric calibration to meet its science goals, particularly for cosmological analyses based on Type Ia supernovae. The Collimated Beam Projector (CBP) has been developed to support this effort by projecting monochromatic point sources of known wavelength and flux directly into the telescope aperture, enabling direct in-situ measurements of the full system throughput. We present initial results demonstrating the CBP's capability to characterize the instrumental response of the Simonyi Telescope and to measure the transmission profiles of LSSTCam broadband filters. In particular, the CBP enables spatially resolved mapping of filter bandpass edge shifts across the focal plane, which can vary by several nanometers as a function of the ray angle of incidence. These early results establish the CBP as a powerful photometric calibration tool and lay the groundwork for continuous throughput monitoring throughout LSST operations. 
\end{abstract}


\section{Introduction}\label{sec:intro} 

As large-scale photometric surveys enter an era of unprecedented statistical precision, correspondingly precise instrumental calibration has become one of the dominant sources of systematic uncertainty. Among the most demanding science cases is the measurement of cosmological parameters from Type Ia supernovae. Results from recent surveys show that this probe still carries among the largest systematic error budgets \cite{abbott2024des,Brout2022}.

The Legacy Survey of Space and Time (LSST), conducted with the Simonyi Telescope at the Vera C. Rubin Observatory, will detect an unprecedented number of Type Ia supernovae over its ten-year baseline \cite{Howlett2017}. As the statistical sample grows, the dominant contribution to the cosmological parameter uncertainty will shift from statistical to systematic, with photometric calibration being one of the most critical \cite{betoule2014}. Meeting the survey science goals therefore requires achieving per-mil photometric precision in the calibration of the instrumental response \cite{stubbs2015}.

Traditional photometric calibration strategies include flat-fielding \cite{marshall2013} and observations of spectrophotometric standard stars such as CALSPEC sources \cite{calspec}. These approaches have historically achieved calibration at the percent level, but are subject to well-known limitations: flat-fielding is sensitive to the non-uniformity of the light source, while standard-star observations are affected by atmospheric transmission, sky background contamination, and the finite accuracy of the reference Spectral energy distribution (SEDs) \cite{bohlin2014}. To address the instrumental-throughput component of this calibration challenge, and in particular the limitations associated with conventional flat-field measurements, the Collimated Beam Projector (CBP) was developed as a dedicated in-dome photometric calibration instrument \cite{ingraham2016}.

More fundamentally, the CALSPEC celestial calibration sources are based on white dwarf model atmospheres. The theoretical white dwarf spectra have exhibited changes over the past decade that exceed one percent, which exceeds our calibration uncertainty budget. An alternative calibration methodolgy \cite{stubbs2006toward} is to use silicon photodiodes as the primary flux calibration reference, explouting the fact that the QE of these sensors can be determined to better than a part per thousand. This reference diode is used to determine the dose of flux, at each wavelength, that is delivered to the pupil of the telescope. We can compare the results using both celestial and terrestrial calibration references to assess consistency. 

The CBP operates by injecting a collimated, monochromatic beam of known flux and wavelength directly into the telescope aperture, producing artificial point-like and monochromatic light source, mimicking the behaviour of a star. This configuration bypasses the atmosphere and eliminates stray-light artifacts associated with diffuse illumination methods, enabling direct throughput measurements of the full optical system with sub-nanometer wavelength accuracy. 

Several generations of the CBP have been deployed and validated on different telescopes. Earlier versions installed on the PanSTARRS and CTIO Blanco telescopes demonstrated the feasibility of the concept at an operational facility \cite{stubbs2007preliminary,stubbs2010precise, coughlin2018ctio}. A more recent implementation on the StarDICE telescope achieved a precision of approximately 0.2\,nm in the determination of filter edge positions, resulting in a contribution below one per-mil to the broadband flux uncertainty when combined with complementary calibration data \cite{souverin_stardice_2024}.

This paper focuses on the Rubin Observatory CBP and its early science results. The CBP operates in conjunction with other in-dome calibration systems, including the flat-screen projector, which is designed for full-pupil illumination and flat-field calibration. Together, these systems provide calibration coverage during both daylight and overcast periods when on-sky data cannot be acquired, and enable controlled-conditions measurements with improved systematic control.

This paper is organized as follows. Section~\ref{section:requirements} describes the CBP design requirements and instrumental configuration. Section~\ref{section:workflow} presents the data acquisition strategy, the analysis workflow, and the datasets used in this study. Section~\ref{section:telescope_reponse} describes how the Rubin instrumental response is estimated from CBP measurements, and Section~\ref{section:filter_throughputs} presents the measurements of the LSST filter throughputs. We conclude in Section~\ref{section:conclusion}.

\section{Requirements and Instrumental Setup}\label{section:requirements}

\subsection{The Collimated Beam Projector concept}

\begin{figure}
    \centering
    \includegraphics[width=0.7\linewidth]{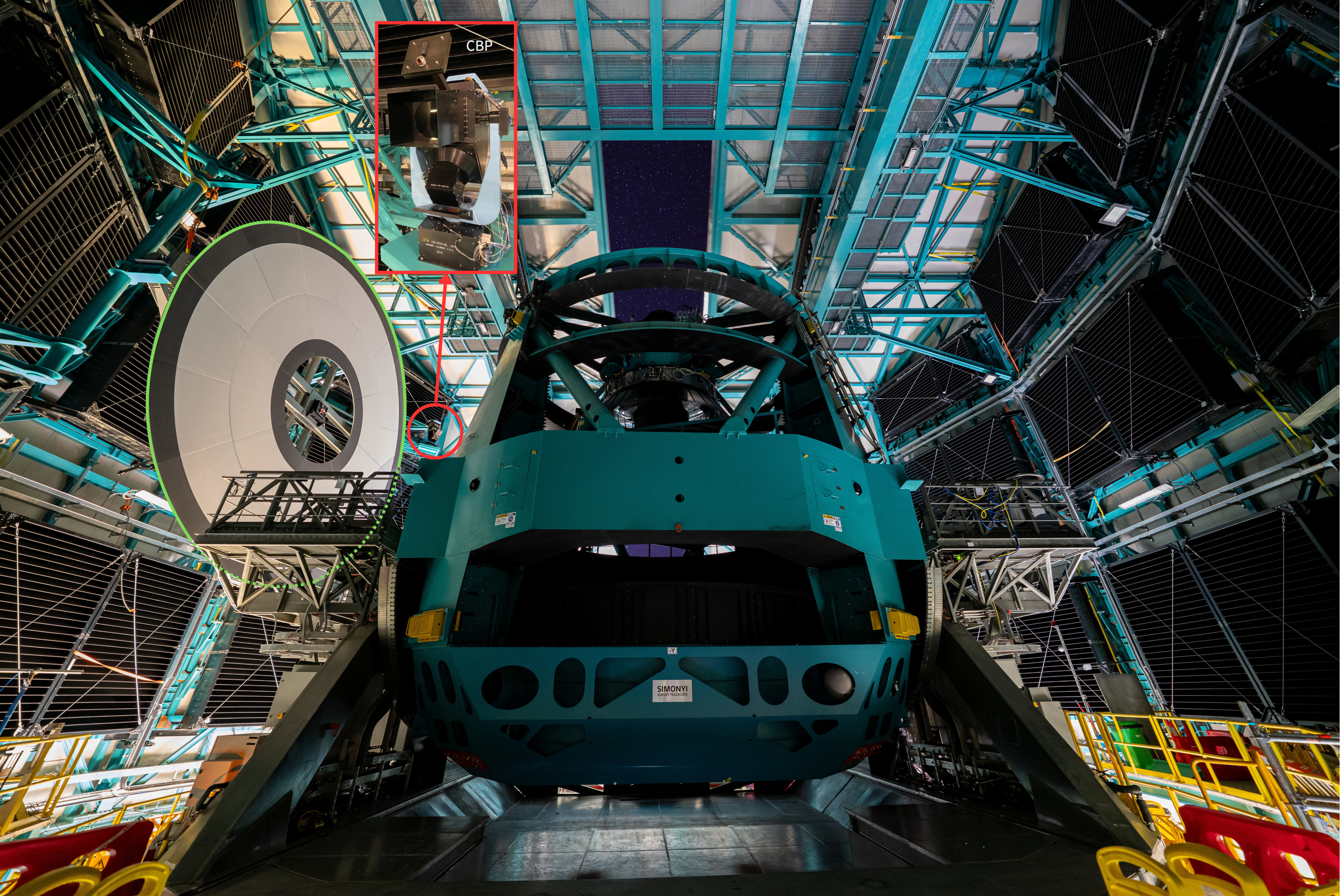}
    \caption{In-dome view of the Simonyi Survey Telescope (center), with the flat-screen projector highlighted in green and the CBP highlighted in red. An enlarged view of the CBP is shown in the red inset.}
    \label{fig:cbp_indome}
\end{figure}

The Rubin CBP is one of the in-dome photometric calibration subsystems designed for the LSST. Its integration within the observatory calibration infrastructure is illustrated in Figure~\ref{fig:cbp_indome}. The instrument comprises three main subsystems: an optical telescope assembly, a tunable monochromatic light source, and a combined photometric and spectroscopic flux monitoring system.

These subsystems enable the projection of a collimated, monochromatic beam into the telescope aperture, reproducing the wavefront geometry of distant point sources with controlled flux and wavelength. Because the CBP beam only partially illuminates the telescope entrance pupil at any given pointing, a set of spatially distinct pointings is required to sample the full pupil and obtain a complete characterization of the instrumental response. This approach avoids the systematic errors associated with diffuse flat-screen illumination, notably stray light and ghosting\footnote{The CBP is not entirely free from ghosting effects; however, the level of contamination is significantly lower than for flat-field illumination, and the resulting ghost images are generally spatially separated from the primary spots.}. In combination with the flat-screen projector and other calibration data products, the target is to achieve per-mil photometric precision for the LSST survey.

\subsection{Design Requirements} 

The Rubin CBP was designed to satisfy a set of requirements ensuring its effectiveness as a photometric calibration reference. The principal requirements are as follows:
\begin{itemize}
    \item \textbf{Wavelength coverage:} The CBP must span the full LSST bandpass, from 300\,nm to 1100\,nm, providing calibration across all photometric filters.
    \item \textbf{Monochromaticity:} The output spectral bandwidth must remain below 1\,nm to accurately simulate monochromatic sources and limit spectral contamination.
    \item \textbf{Beam collimation:} The output wavefront must be sufficiently collimated to mimic the parallel ray geometry of distant astrophysical sources at the telescope entrance pupil.
    \item \textbf{Mask versatility:} A set of interchangeable focal-plane masks must be available to produce different illumination patterns suited to specific calibration objectives, including ghost and crosstalk characterization, throughput measurements, and CBP self-calibration.
    \item \textbf{Beam uniformity:} The intensity profile of the output beam must be spatially uniform to ensure consistent and reproducible calibration results.
    \item \textbf{Focal-plane coverage:} A single CBP pointing must simultaneously illuminate all CCDs of the LSSTCam focal plane, enabling efficient camera-wide characterization.
    \item \textbf{Pupil Pivoting:} The CBP must pivot about its pupil. This ensures that we can project beams that emerge from the CBP at different angles along the same path into the pupil.
    \item \textbf{Temporal stability:} The CBP must deliver stable and reproducible output flux over the 10 years LSST operations, enabling consistent cross-epoch calibration.
    \item \textbf{Field of View:} The CBP must generate high-quality PSFs from masks that span a field of view that matches the Rubin 3.5 degree field. The Schmidt configuration of the Rubin CBP achieves this goal.
    \item \textbf{Observatory integration:} The system must be compatible with the Rubin Observatory control architecture to allow for remote operation and automated monitoring.
\end{itemize}

\subsection{Instrumental setup}

\begin{figure}[h]
    \centering
    \includegraphics[width=0.4\textwidth,valign=c]{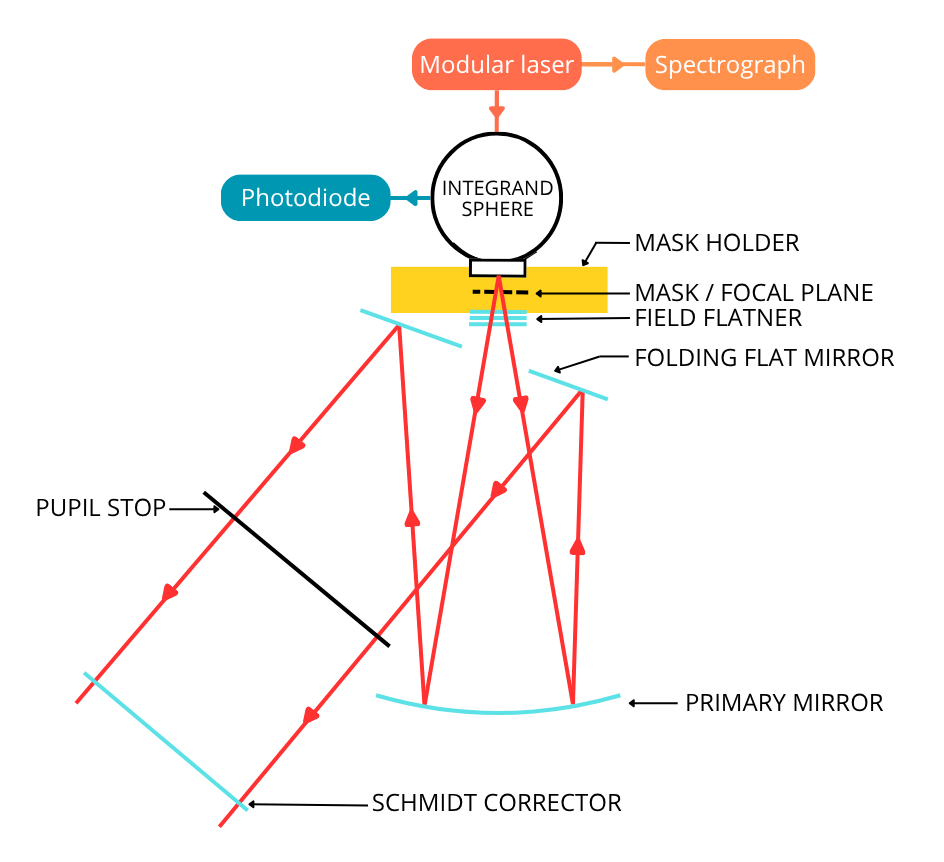}
    \includegraphics[width=0.45\textwidth,valign=c]{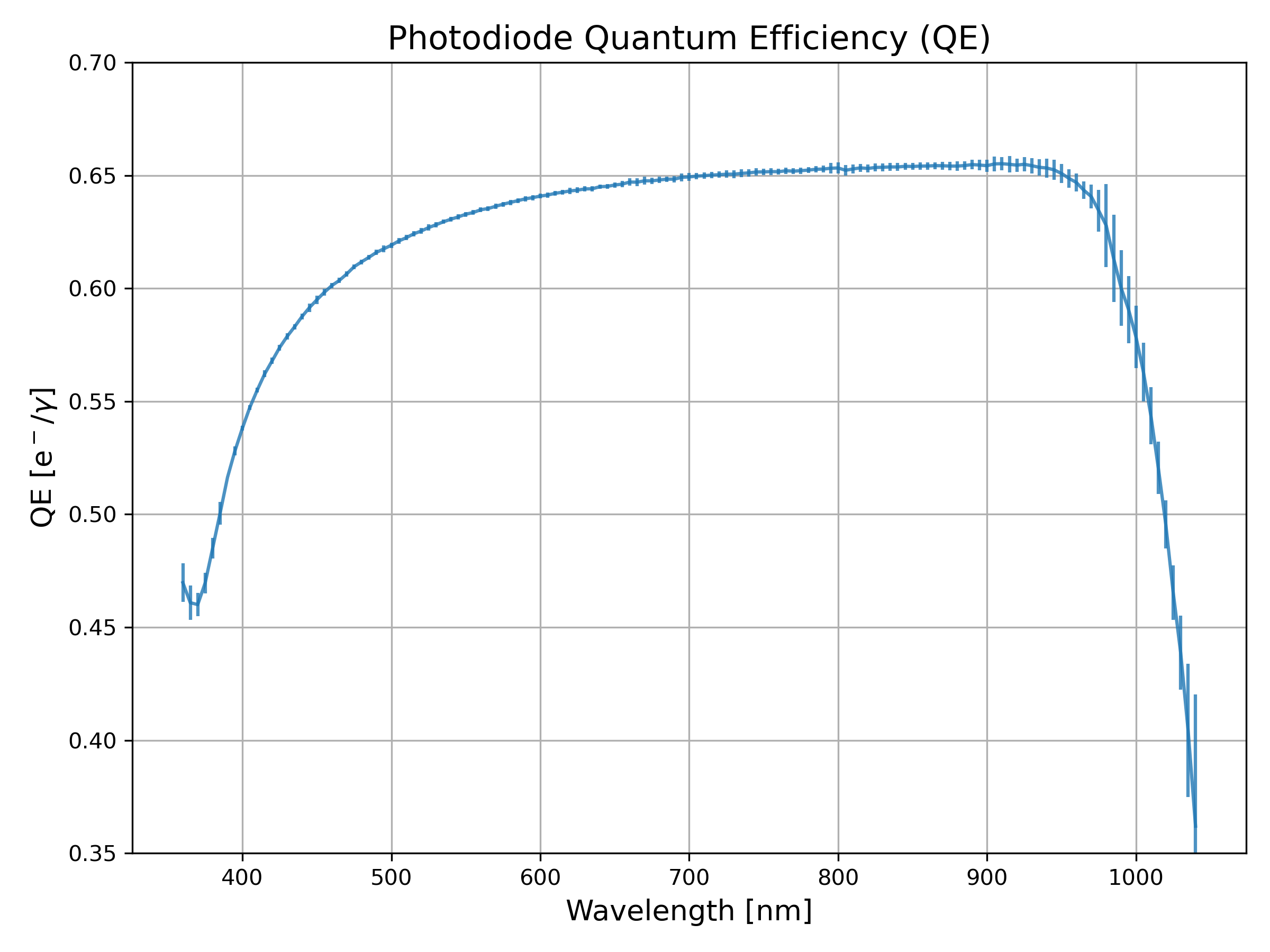}
    \caption{Left: Schematic of the CBP optical system. Right: Quantum efficiency curves of the photodiode used in the CBP monitoring system. Error bars are multiplied by ten for visibility.}
    \label{fig:cbp_layout_and_pd_qe}
\end{figure}

The Rubin CBP is designed to partially illuminate the entrance aperture of the telescope, which has an 8.4-meter primary mirror, while simultaneously producing spots on all CCDs of the LSSTCam focal plane. A schematic of the CBP optical system is shown in Figure~\ref{fig:cbp_layout_and_pd_qe}. The optical train consists of a primary concave mirror that collimates the diverging beam onto a flat folding mirror coincident with the CBP exit pupil. The primary mirror position can be adjusted by several millimeters (in 5\,\textmu m increments) to fine-tune the collimation of the output beam. A Schmidt corrector and three field-correction lenses minimize wavefront aberrations and suppress residual scattered light.

The CBP has a focal length of 0.635\,m and an aperture of 0.241\,m (f/2.6), corresponding to a linear magnification factor of 16 between the mask plane and the LSSTCam focal plane. Consequently, a 100\,\textmu m pinhole mask at the CBP focal point produces a spot of approximately 1.6\,mm (160 pixels) on the detector. The optical assembly is mounted on an alt-azimuth mount, enabling the CBP beam to be directed at different positions on the primary mirror. The instrument is installed on a dedicated platform approximately 10\,m above the dome floor, as shown in Figure~\ref{fig:cbp_indome}.

Surface brightness uniformity at the mask plane is achieved using a Labsphere 3P-GPS-060-SF integrating sphere equipped with a Spectraflect interior coating, which provides high-reflectivity Lambertian diffuse reflectance over a broad wavelength range. The integrating sphere also eliminates optical memory effects from the upstream illumination path. The calibration diode monitors the internal flux density within the integrating sphere, and this is used to determine the delivered photon dose. At the largest exit port of the integrating sphere, coincident with the focal plane of the CBP telescope, a mask wheel provides a selection of illumination masks for different calibration configurations, including ghost and crosstalk characterization, throughput mapping, and CBP self-calibration.

The light source is an Ekspla NT-242 tunable pulsed laser coupled to the integrating sphere via an optical fiber. The laser is based on a Q-switched Nd:YAG pump architecture, providing narrow-band tunable illumination over the range 300–2300\,nm. For CBP operations, the laser is operated in burst mode, where each burst consists of a user-defined sequence of individual laser pulses emitted over a short time interval. The delivered photon flux can therefore be adjusted by changing either the number of bursts or the number of pulses per burst. This configuration increases the effective photon statistics per acquisition while preserving the temporal structure of individual pulses, which facilitates synchronization with external triggers and reduces readout-dominated noise. The spectral bandwidth of the laser ranges from 0.05\,nm in the ultraviolet to approximately 1\,nm at wavelengths above 710\,nm.

Flux monitoring is performed with a Hamamatsu silicon photodiode coupled to one of the output ports of the integrating sphere. The photodiode signal is read out by a Keysight B2985B electrometer, which records the accumulated charge and provides a measurement proportional to the photon flux delivered by the sphere. The photodiode has an active area of 100\,mm$^2$ and is calibrated against NIST standards; its quantum efficiency (QE) is known to better than the part per thousand level. The wavelength-dependent QE curve is shown in Figure~\ref{fig:cbp_layout_and_pd_qe}.

Spectroscopic monitoring is provided by a pair of Avantes fiber spectrographs covering 320–1140\,nm with a spectral resolution of 0.75–0.85\,nm depending on wavelength. This system is intended to verify spectral purity and flux stability throughout a calibration sequence. However, the spectrographic system was not operational at the time of the data campaigns analyzed in this work; we therefore treat the laser output wavelength as equal to the commanded value and assume a sufficiently narrow linewidth.

\section{Dataset Description, Workflow, and Data Analysis}\label{section:workflow}

\subsection{Taking CBP Data}

In this work, we use the single-spot-per-CCD mask configuration, where illuminated pinholes in the CBP mask are imaged onto the LSSTCam focal plane, producing one spot on each detector. This choice achieves adequate sampling of the LSSTCam focal plane while maintaining a high signal-to-noise ratio per spot and minimizing contamination from ghost images. To resolve the positional degeneracy between spots and facilitate their identification, the central CCD spot is offset by 0.5\,mm along the negative $y$-axis of the mask plate. This offset translates to a displacement of approximately 10\,mm on the focal plane, providing a straightforward visual check of mask alignment in the acquired images.

Accurate co-pointing of the CBP and the Rubin Telescope is essential for efficient observations, yet challenging given the CBP's fixed position within the dome. To ensure that spots are centered on the intended CCDs, the optical axes of both instruments must be precisely aligned. We developed a co-pointing model\footnote{\url{https://pipelines.lsst.io/modules/lsst.cbp/configuration.html\#lsst-cbp-configuration}} based on laser tracker survey data, which computes the required CBP and Telescope Mount Assembly (TMA) altitude-azimuth coordinates for a specified target position on the pupil or focal plane.

The wavelength sampling for each CBP campaign is determined by the calibration objective and the time available. For no-filter (None filter herafter) telescope response measurements, the full CCD wavelength range is sampled in 20\,nm steps, providing adequate spectral coverage within a reasonable acquisition time. Filter-edge regions require finer sampling (typically 2\,nm) to accurately characterize the steep transmission variations at bandpass boundaries; a coarser step (5\,nm) is used across the filter plateaus.

The delivered photon flux is controlled by adjusting the number of laser bursts and pulses per burst. These parameters are optimized to maximize the signal-to-noise ratio on the CCDs while avoiding saturation of either the photodiode or the camera, accounting for the wavelength-dependent CBP and telescope transmission profiles. The optimal parameter values were determined empirically through commissioning campaigns. In some cases, residual saturation was encountered on the photodiode or on individual CCDs, requiring targeted fine-tuning.

\subsection{Raw Data Products}

Two complementary data streams are required to extract the CBP calibration signal: the incident flux measured by the monitoring photodiode, and the detected signal recorded by the LSSTCam CCDs.

The incident flux from the CBP is measured by the Hamamatsu photodiode positioned at one of the exit ports of the integrating sphere. The electrometer records the accumulated charge as a function of time, from which the total number of photoelectrons collected during an exposure is computed. This quantity is directly proportional to the total photon flux directed toward the telescope.

\begin{figure}
    \centering
    \includegraphics[width=0.4\linewidth,valign=c]{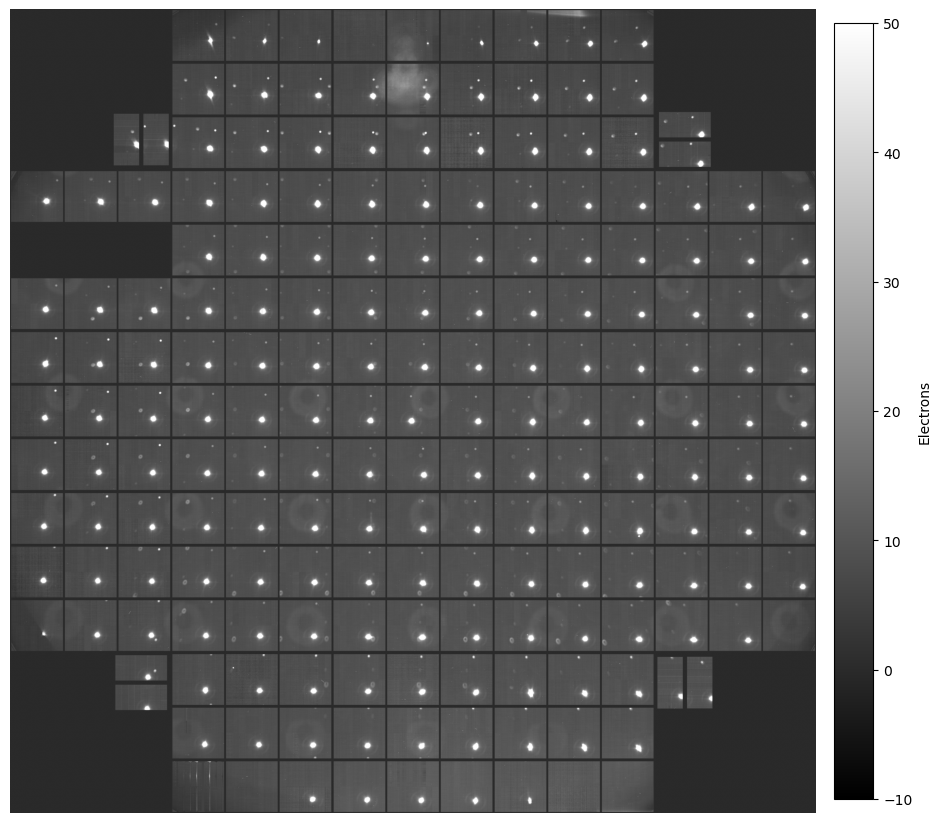}
    \includegraphics[width=0.5\linewidth,valign=c]{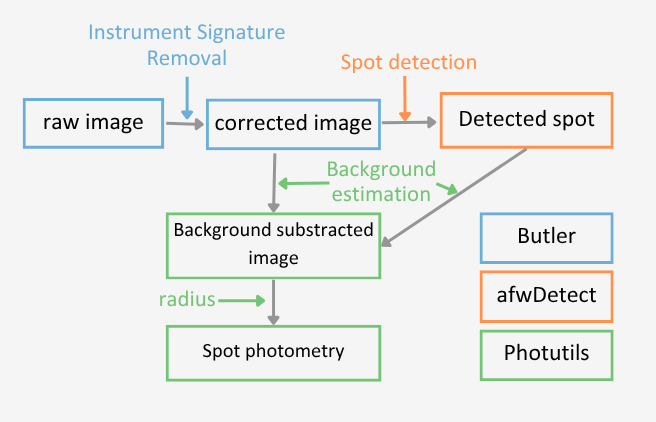}
    \caption{Left: LSSTCam focal-plane image of CBP spots after instrument signature removal, at a single wavelength. Right: Schematic diagram summarizing the main steps of the photometric extraction workflow.}
    \label{fig:focal_plane_and_workflow}
\end{figure}

The signal detected by LSSTCam is spatially distributed across the focal plane according to the illumination pattern imposed by the mask. Left pannel of Figure~\ref{fig:focal_plane_and_workflow} shows a representative focal-plane image acquired at a single wavelength. The faint secondary spots and donut-shaped features visible in the image are ghost reflections from surfaces within the optical path. These ghosts can originate both from LSSTCam and CBP optics. The brightest ghost features contain less than one percent of the flux of the corresponding primary spot. In the datasets analyzed here, they are located outside the photometric aperture of the parent spot and do not produce any significant overlap with the apertures used for neighboring primary spots. Their contribution to the extracted photometry is therefore neglected. Our analysis extracts the photometric signal from the primary spot on each individual CCD with aperture photometry after background substraction.

\subsection{Analysis Workflow}

The processing steps described below are summarized in the diagram of Figure~\ref{fig:focal_plane_and_workflow} right and are implemented within the Rubin Data Management (DM) software framework.

Instrument signature removal is applied using a custom configuration adapted for CBP data. Critically, flat-field correction is intentionally omitted in order to preserve the CCD-to-CCD response variations that the CBP is designed to measure. Brighter-fatter\cite{Broughton_2024} corrections and pixel interpolation are also disabled; the latter is particularly important when processing partially saturated images, where interpolation would corrupt the integrated charge measurement.

Spot detection is performed using the DM built-in algorithm, which applies an image-adaptive threshold based on the local mean and standard deviation of pixel values. A preliminary aperture photometry step provides an initial flux estimate, which is used to identify the brightest spots as positional references. The final flux extraction is then performed using forced aperture photometry at the median reference spot position on each detector.

A dedicated photometry pipeline, developed to be compatible with the Rubin DM software environment, is applied to each detector at each wavelength. Background subtraction precedes photometry: the spot region is masked with a 1200-pixel aperture centered on the fitted position, and a two-dimensional background model is estimated using 400-pixel sub-boxes to capture low-spatial-frequency gradients across the image. The fitted background is subtracted from the instrument-signature-removed image. Background levels are in general negligible compared to the spot flux, as illustrated by the spot profiles shown in Figure~\ref{fig:spot_profiles}, so the adopted background model does not need to be highly detailed.

Aperture photometry is then performed with a 600-pixel aperture radius, yielding the CCD signal $\rm S_{\rm CCD}$ in units of electrons.

On the photometric monitoring side, the photodiode signal $\rm S_{\rm PD}$ (in coulomb) is extracted for each wavelength by computing the peak-to-valley variation of the accumulated charge over the acquisition window.

\begin{figure}
    \centering
    \includegraphics[width=0.49\linewidth,valign=c]{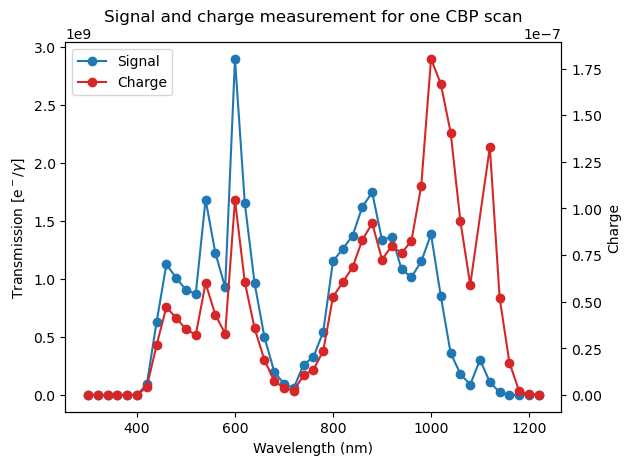}
    \includegraphics[width=0.46\linewidth,valign=c]{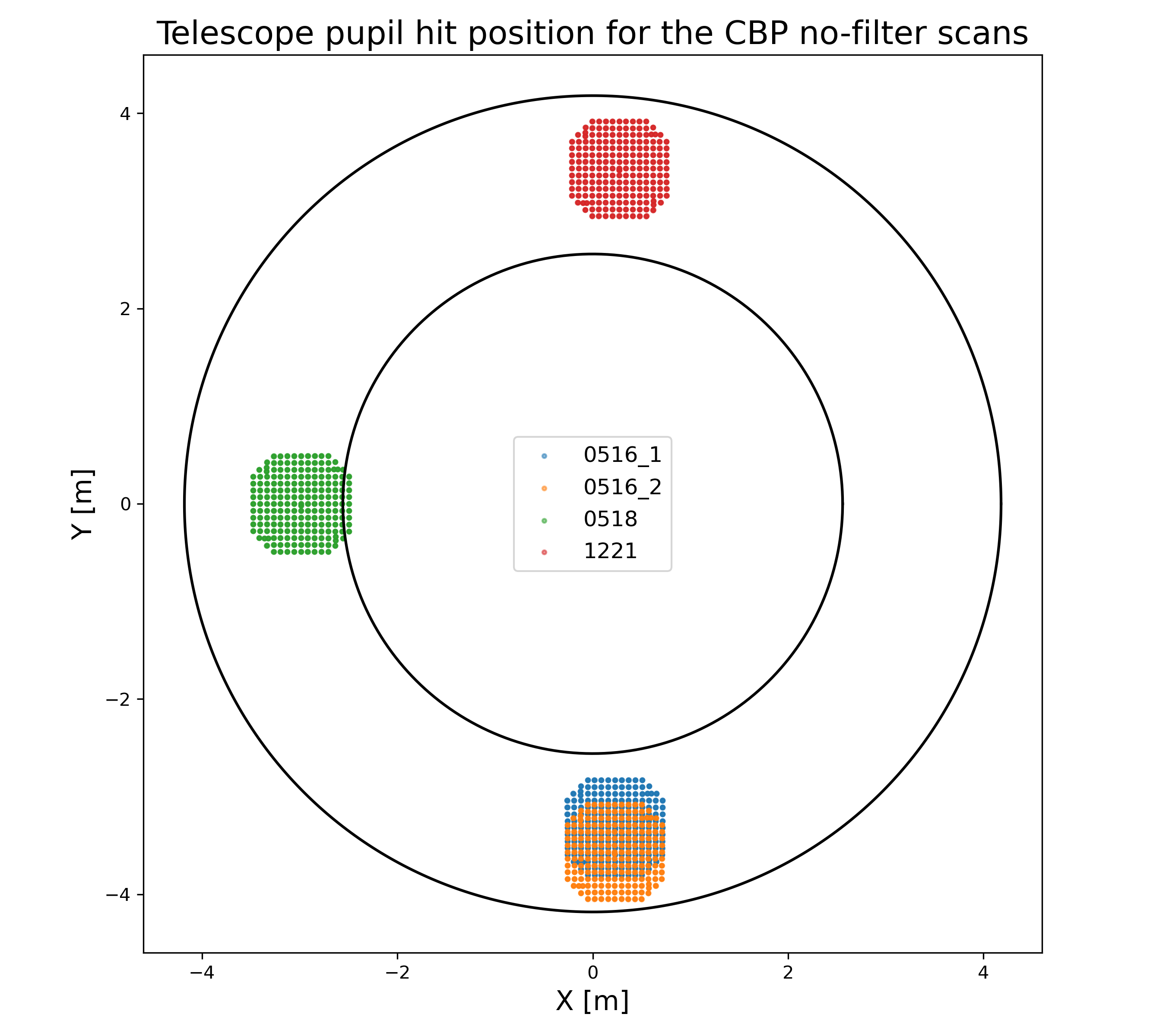}
    \caption{Left: Photodiode charge and CCD signal as a function of wavelength for a single CBP campaign. Right: Map of the primary mirror pupil showing the beam footprint positions for each no filter campaign.}
    \label{fig:charge_and_signal_one_campaign_and_pupil_layout}
\end{figure}

Figure~\ref{fig:charge_and_signal_one_campaign_and_pupil_layout} shows both the photodiode signal and the CCD signal as a function of wavelength for a single campaign. The elevated photon flux at wavelengths beyond 900\,nm reflects the steep drop in telescope transmission in that region, which requires the CBP to deliver approximately twice the photon flux relative to the bluer portion of the bandpass to achieve comparable signal levels.

\subsection{CBP Campaigns}

Following LSSTCam installation and commissioning, several CBP datasets were acquired. Table~\ref{table:datasets} summarizes the campaigns analyzed in this work. For each campaign, the wavelength range and sampling interval were adapted to the specific science objective: filter surveys employ finer wavelength steps near the bandpass edges to accurately characterize the transmission fronts, while coarser steps are used across the plateaus.

Different regions of the primary mirror were sampled across campaigns, as illustrated in Figure~\ref{fig:charge_and_signal_one_campaign_and_pupil_layout}. Spatially sampling the full pupil is essential to characterize mirror-dependent effects on the instrumental response, such as spatial variations in reflectivity. Each mirror pointing produces a distinct vignetting pattern on the focal plane, which must be corrected. This aspect is addressed in Section~\ref{section:telescope_reponse}.

\begin{table}[h]
    \centering
    \begin{tabular}{|c|c|c|c|c|}
        \hline
        Filter & Date & Wavelength ranges (nm) & Wavelength steps (nm) & Mirror position \\
        \hline
        None & 20250516 & [300, 1220] & 20 & inner bottom \\
        None & 20250516 & [300, 1220] & 20 & outer bottom \\
        None & 20250518 & [300, 1220] & 20 & inner left \\
        None & 20251221 & [300, 1220] & 20 & outer top \\
        g & 20251221 & [385, 415] $\cup$ ]415, 530] $\cup$ ]530, 570] & 2, 5, 2 & outer bottom \\
        r & 20250518 & [535, 565] $\cup$ ]565, 675] $\cup$ ]675, 705] & 2, 5, 2 & outer bottom \\
        r & 20250518 & [535, 565] $\cup$ ]565, 675] $\cup$ ]675, 705] & 2, 5, 2 & inner bottom \\
        i & 20251221 & [665, 705] $\cup$ ]705, 800] $\cup$ ]800, 840] & 2, 5, 2 & outer bottom \\
        z & 20250517 & [800, 830] $\cup$ ]830, 910] $\cup$ ]910, 940] & 2, 5, 2 & outer top \\
        y & 20250518 & [905, 945] $\cup$ ]945, 1120] & 2, 5 & inner left \\
        \hline
    \end{tabular}
    \vspace{6pt}
    \caption{Summary of CBP datasets used for telescope and filter throughput measurements; for multiple wavelength ranges, the listed wavelength steps apply in the same order.}
    \label{table:datasets}
\end{table}

\section{Telescope Response Measurement}\label{section:telescope_reponse}

The telescope throughput is determined by comparing the photon flux incident at the entrance pupil, measured by the CBP monitoring photodiode, with the number of photoelectrons detected by the LSSTCam CCDs. The analysis presented in this section uses no-filter (None) datasets.
\vspace{1cm}

The total instrumental response of the Rubin Telescope, expressed in detected photoelectrons per incident photon, can be written as:
\begin{equation}
    R_{\mathrm{Rubin}}(\lambda, i) = T_{\mathrm{mirrors}}(\lambda) \times T_{\mathrm{lenses}}(\lambda) \times T_{\mathrm{filter}}(\lambda) \times \mathrm{QE}_{\mathrm{CCD}}(\lambda, i)
\end{equation} 
where $T_{\mathrm{mirrors}}$, $T_{\mathrm{lenses}}$, and $T_{\mathrm{filter}}$ denote the wavelength-dependent transmissions of the mirrors, lenses, and filter, respectively; $\mathrm{QE}_{\mathrm{CCD}}(\lambda, i)$ is the quantum efficiency of the $i$-th CCD at wavelength $\lambda$; and $i$ indexes the detector position on the focal plane. The filter term is absent in the no-filter configuration.

Using the CBP, the partial measured telescope response is given by:

\begin{equation}
    R(\lambda, \mathrm{i}) = \frac{\mathrm{S_{CCD}}}{S_{\mathrm{CBP}}} = S_{\mathrm{CCD}}(\lambda, i) \times \frac{e \times \mathrm{QE}_{\mathrm{PD}}(\lambda)}{S_{\mathrm{PD}}(\lambda)\times T_{\mathrm{CBP}}(\lambda)}
    \label{equation:telescope_response}
\end{equation}
where $S_{\rm CCD}$ is the aperture photometry signal from a given CCD (in electrons), $S_{\rm PD}$ is the charge collected by the monitoring photodiode (in coulomb), $T_{\rm CBP}$ is the CBP transmission (in electrons per photon), $e$ is the elementary charge, and $\mathrm{QE}_{\mathrm{PD}}$ is the photodiode quantum efficiency.

\begin{figure}
    \centering
    \includegraphics[width=0.47\linewidth]{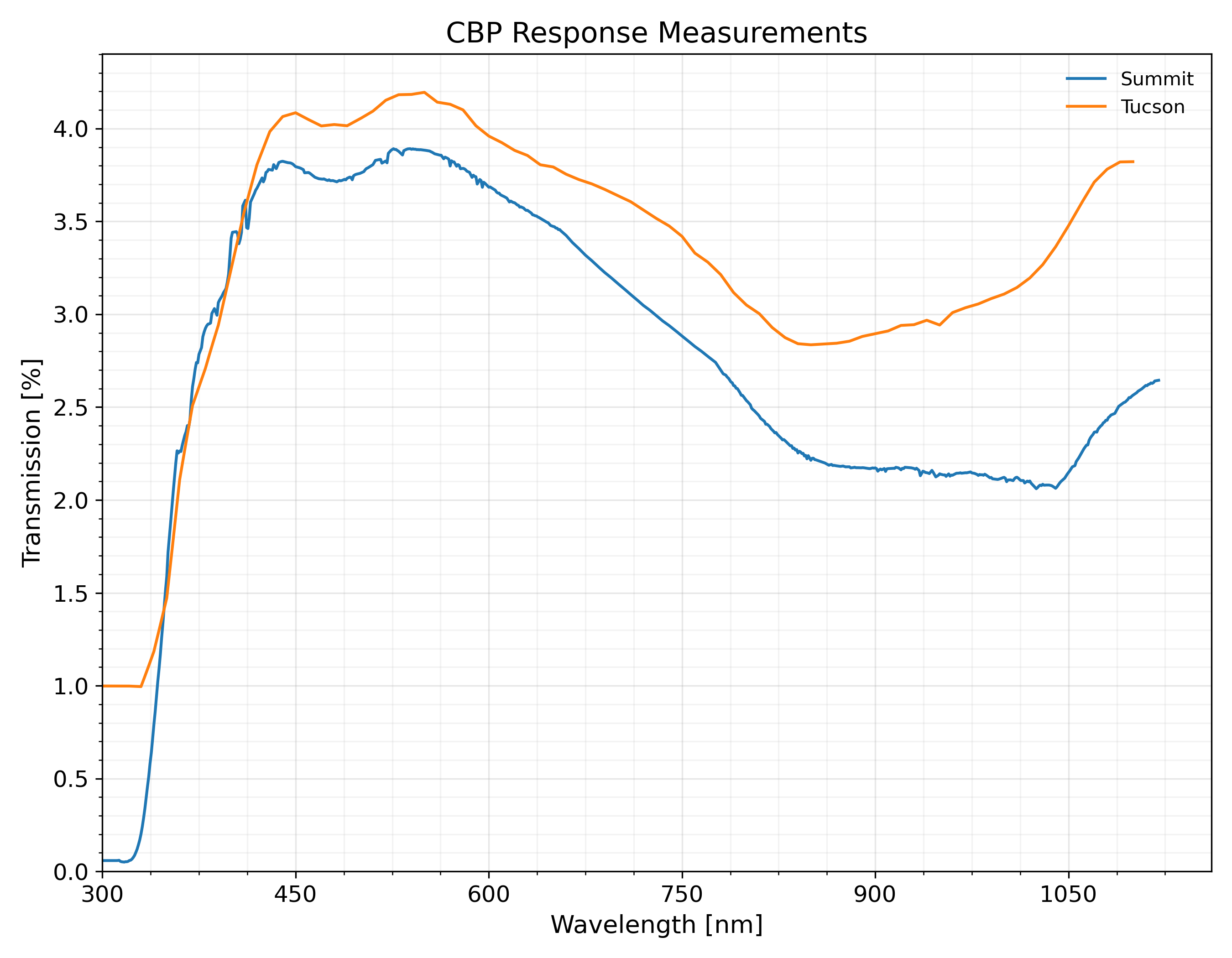}
    \includegraphics[width=0.47\linewidth]{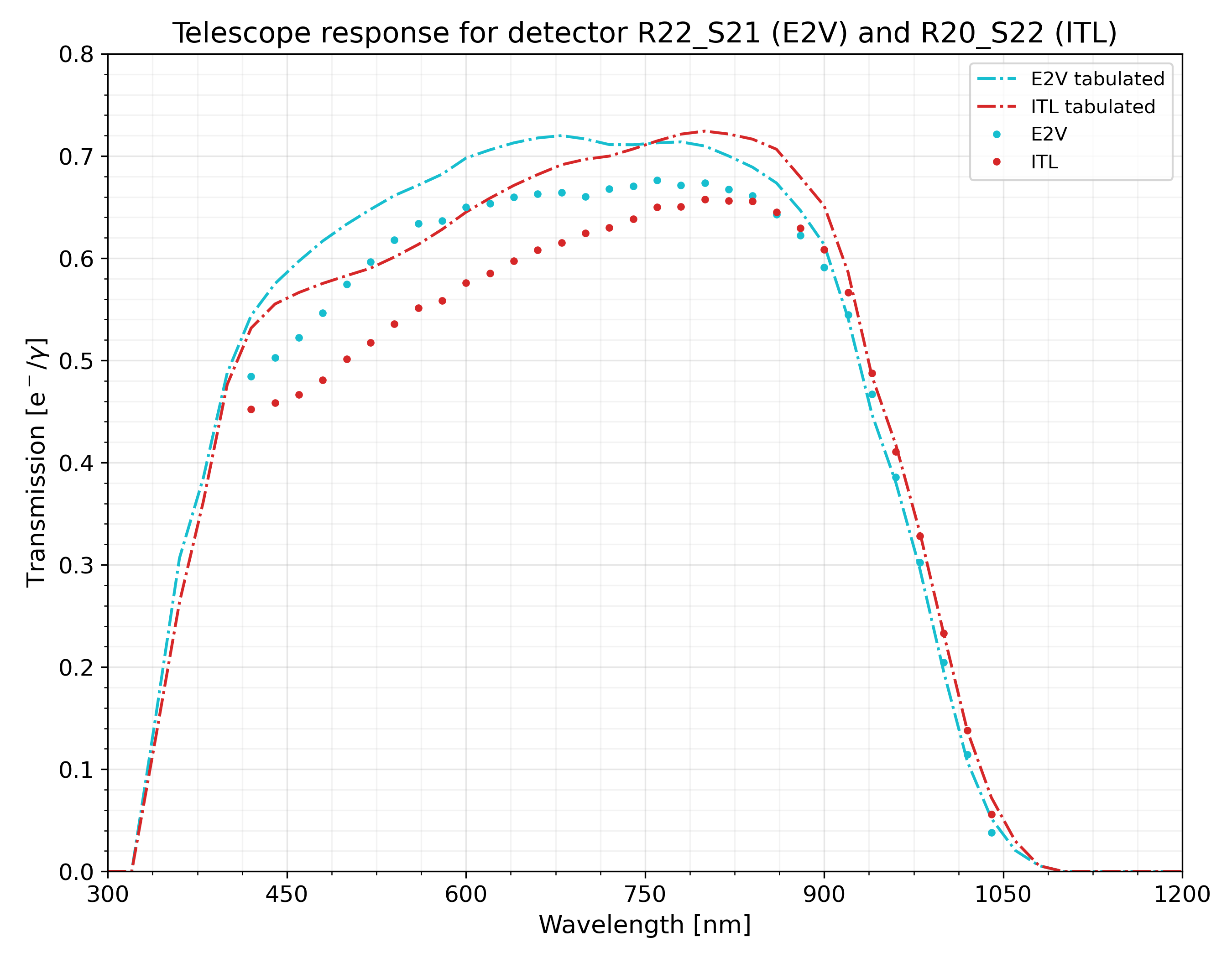}
    \caption{Left: CBP transmission curves from two independent measurements. The blue curve was obtained at the summit following delivery, while the orange curve was measured in the Tucson laboratory prior to shipment. Right: Resulting telescope response curves for two detectors, illustrating the sensitivity of the calibration to the adopted CBP transmission.}
    \label{fig:cbp_response}
\end{figure}

Since the monitoring diode monitors flux at the input of the collimating optic, precise knowledge of the CBP transmission $T_{\rm CBP}$ is a prerequisite for determining the relative telescope plus instrumental response. The CBP was characterized under controlled laboratory conditions in Tucson prior to shipment to the observatory. Following reassembly at the summit, a second transmission measurement was performed using a nominally identical calibration setup. The two resulting curves, shown in Figure~\ref{fig:cbp_response}, exhibit significant discrepancies, particularly at wavelengths above 700\,nm, where the two characterizations diverge by more than 15\%. The origin of this discrepancy is attributed to suboptimal laboratory conditions during the summit measurement. Consequently, the relative telescope response derived from these data carries a systematic uncertainty that currently precludes a reliable precision calibration\footnote{The CBP transmission includes an implicit geometric correction factor of $(1/3)^2$, accounting for the photodiode aperture radius being three times larger than the beam exit-port radius. As a result, only $1/9$ of the photodiode area subtends the beam solid angle. The displayed curves doesn't include this factor}. 

\vspace{3cm}

Although the CBP throughput cannot currently be relied upon, the CBP data can nevertheless be used to characterize relative variations across the focal plane, which are independent of the CBP transmission. For each CCD type (t $\in {\mathrm{e2v}, \mathrm{ITL}}$), we define the relative quantum efficiency (RQE) of detector $i$ as the ratio of its signal to the mean signal over all detectors of the same type:

\begin{equation}
    \mathrm{RQE}(i,\lambda) = \frac{S_{\mathrm{CCD}}(i,\lambda)}{\overline{S_{\mathrm{CCD},\mathrm{t}}}(\lambda)},
\end{equation}

At a given CBP pointing, CCDs located near the edge of the focal plane receive less flux than central ones due to vignetting by the telescope optics. For a full-pupil illumination, vignetting correction models exist and are routinely applied in flat-field analysis. However, because the CBP only partially illuminates the pupil, the resulting vignetting pattern depends on the CBP pointing and is more complex. An equivalent correction model has not yet been developed for the partial-pupil illumination case.

\begin{figure}[t]
    \centering
    \includegraphics[width=.6\linewidth]{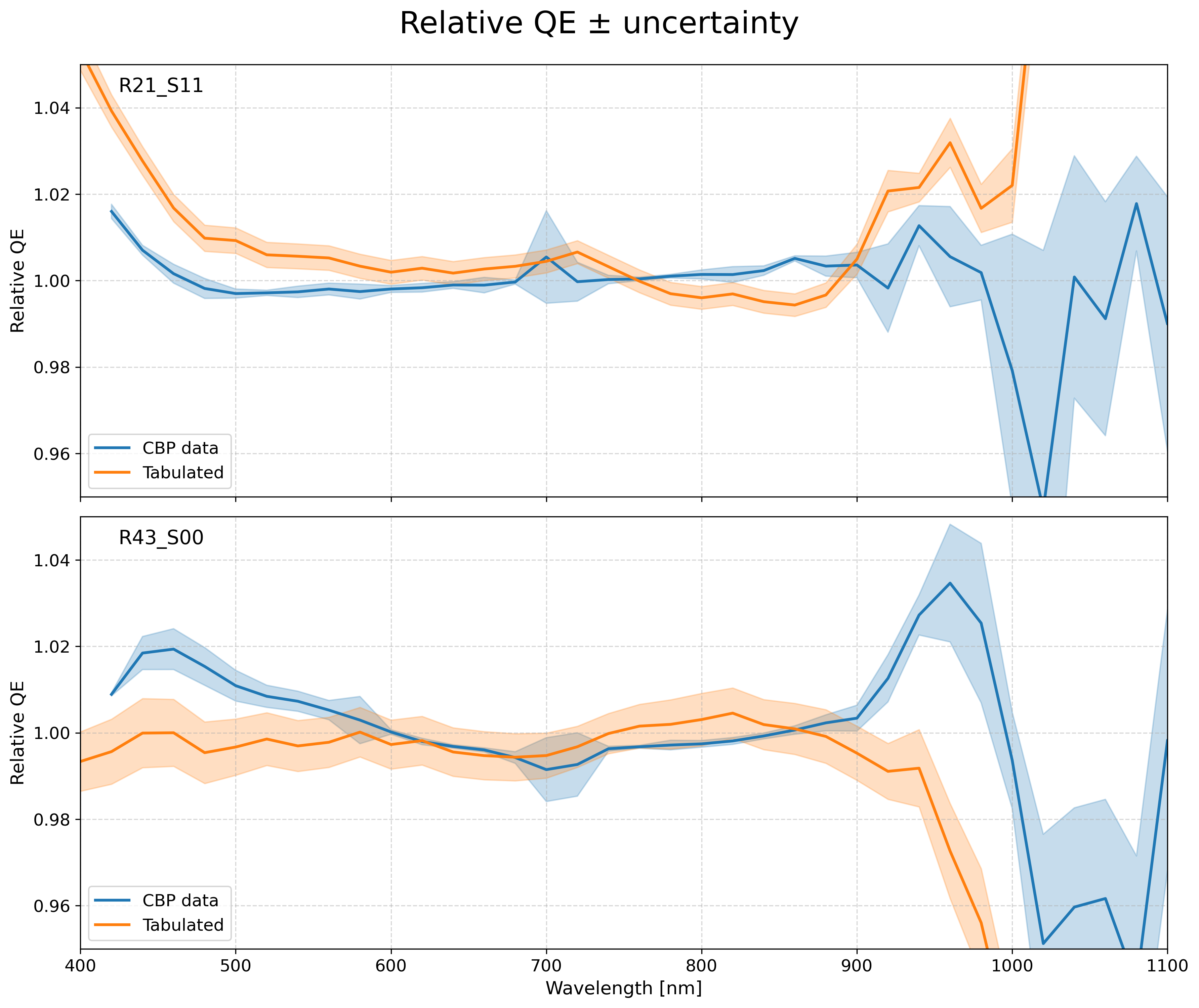}
    \caption{CBP-measured (blue) and tabulated (orange) relative quantum efficiency for two LSSTCam detectors. Top: e2v CCD. Bottom: ITL CCD.}
    \label{fig:relative_qe}
\end{figure}

To mitigate the vignetting bias on the RQE, we apply a per-detector normalization factor defined as the ratio between each detector's RQE and the mean RQE over a wavelength range with high signal-to-noise ratio, chosen here as [540,\,640]\,nm $\cup$ [760,\,860]\,nm. This normalization removes the average vignetting offset, at the expense of absorbing genuine inter-detector QE differences in the reference bandpass.

The tabulated and CBP-measured normalized RQE curves for one detector of each family are shown in Figure~\ref{fig:relative_qe}. The tabulated values come from uniform focal plane illumination measurement operated under laboratory conditions during the assembly of the camera. The tabulated uncertainty represents the amplifier-to-amplifier dispersion within the same CCD as provided by the same laboratory measurement. For the CBP measurement, we estimate the uncertainty from the spread between two comparable datasets acquired with different pupil illuminations, as:
\begin{equation}
    \sigma_{\mathrm{RQE}} = \max\left(\frac{\mathrm{RQE}_{\rm inner\text{-}bottom}}{\mathrm{RQE}_{\rm inner\text{-}left}},\ \frac{\mathrm{RQE}_{\rm outer\text{-}bottom}}{\mathrm{RQE}_{\rm inner\text{-}left}}\right).
\end{equation}
This uncertainty is built to encompass both the reproducibility of the CBP measurements and variabilities from configuration changes.

In Figure~\ref{fig:relative_qe}, a systematic offset at the percent level between the CBP-measured and tabulated RQE curves is observed across the 400–860\,nm range. This discrepancy is noteworthy for LSST photometric calibration, as it suggests that laboratory-QE data may be insufficient for the per-mil precision required by the survey, particularly for the Forward Global Calibration Method (FGCM).

\begin{figure}
    \centering
    \includegraphics[width=.8\linewidth]{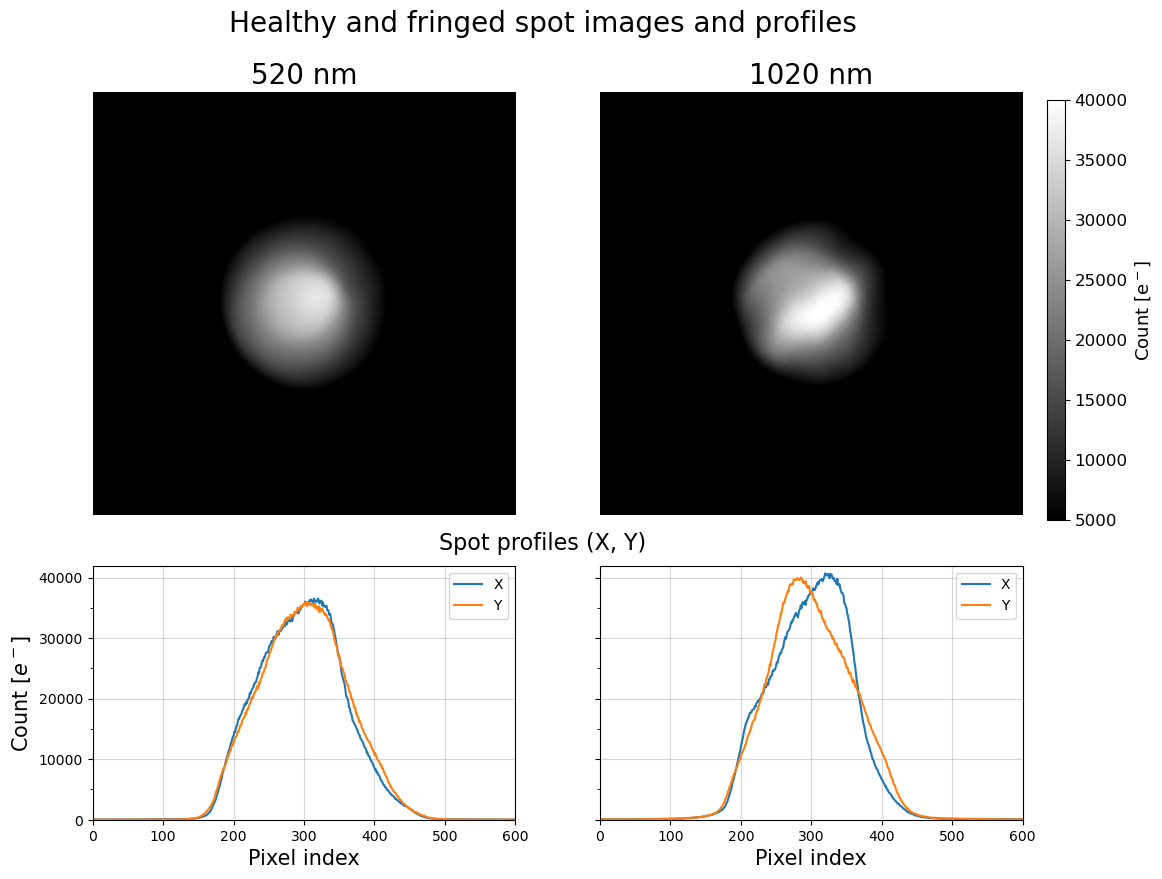}
    \caption{CBP spot images at two wavelengths on the same CCD, illustrating a well-behaved spot at high SNR (top-left) and a fringing-affected spot at near-infrared wavelengths (top-right), with corresponding profiles in the serial (X) and parallel (Y) directions.}
    \label{fig:spot_profiles}
\end{figure}

The RQE uncertainty increases and the curve loses its smoothness at wavelengths beyond 900\,nm. This behavior is attributable to infrared fringing causing the CBP spot profile to be distorted and exhibits flux redistribution between the core and wings that varies between the serial and parallel directions of the CCD, as shown in Figure~\ref{fig:spot_profiles}. Since the fringing pattern differs from detector to detector, it introduces differential response variations of up to 10\% in the relative QE at these wavelengths. Modeling and correcting for fringing will be required to achieve accurate photometric calibration in the $z$ and $y$ bands. A future finer scan in wavelength in this area is mandatory to accurately measure this effect and correct it.

\section{Toward LSST Filter Throughput Measurements}\label{section:filter_throughputs}

The CBP-measured telescope response can be used to derive the filter transmission profiles. The filter throughput is obtained by taking the ratio of the Rubin telescope response measured with and without the filter in the beam. Under the assumption that both measurements are performed with identical pupil pointings, optical transmission terms common to both (mirrors, lenses, detector QE, and vignetting) cancel in the ratio:
\begin{equation}
    R_{\rm filter}(\lambda, i) = \frac{S_{\rm CCD,\,filter}(\lambda, i) \cdot S_{\rm PD,\,no\,filter}(\lambda)}{S_{\rm CCD,\,no\,filter}(\lambda, i) \cdot S_{\rm PD,\,filter}(\lambda)}.
\end{equation}
This ratio is also independent of the CBP transmission and photodiode QE, so it provides a self-contained measurement of the filter throughput that does not require an accurate characterization of the CBP instrument.

Since the filter and no-filter datasets do not necessarily cover the same wavelength grid (see Table~\ref{table:datasets}), the ratio $S_{\rm CCD}/S_{\rm PD}$ is evaluated as a function of wavelength for each dataset individually and interpolated with a cubic spline prior to computing the ratio. The interpolation is performed on the ratio rather than on the individual signals to avoid numerical artifacts: the photodiode signal can vary by orders of magnitude over small wavelength intervals, making interpolated values unreliable for the individual quantities.

\begin{figure}
    \centering
    \includegraphics[width=0.8\linewidth]{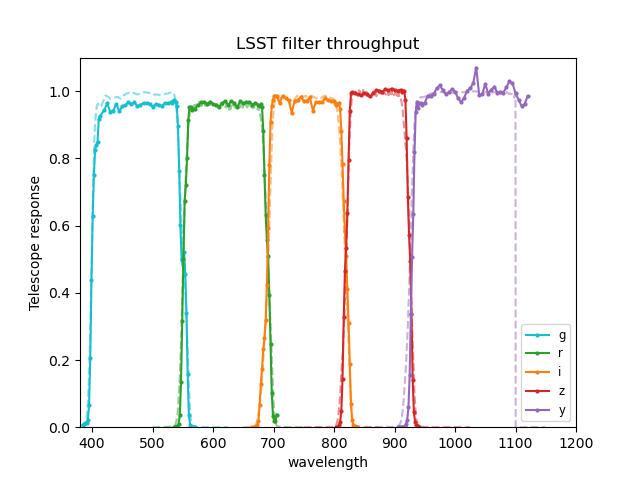}
    \caption{CBP-measured (solid lines) and manufacturer-tabulated (dashed lines) LSST filter throughputs for a single detector.}
    \label{fig:throughputs_and_blueshift}
\end{figure}

Figure~\ref{fig:throughputs_and_blueshift} shows the measured throughput for five of the six LSST filters, derived from a single pupil illumination. Bandpass edges are characterized using 2\,nm wavelength steps, while 5\,nm sampling is used in the high-transmission regions. The measured throughput curves agree well with the tabulated data across these regions, with the exception of the g filter, which shows a 3\% discrepancy. In another hand, the filter fronts are not directly compared here because they would require a sampling of the full telescope pupil through multiple CBP pointings (pupil stitching), which has not yet been performed for all filters.

Oscillatory features are visible in the red-end of the $z$ filter plateau and across the $y$ filter, arising from infrared fringing of the CBP spot (Section~\ref{section:telescope_reponse}). Because the fringing pattern is wavelength-dependent and cannot be captured by the low resolution None filter dataset, the cubic spline interpolation introduces residual errors in the throughput ratio at these wavelengths, producing transmission values that deviate from the tabulated data and in some cases exceed unity. Future measurements with finer wavelength sampling and multiple pupil pointings should reduce this effect, as averaging over different incidence angles is expected to smooth out the fringing contribution.

\begin{figure}[h]
    \centering
    \includegraphics[width=0.47\linewidth]{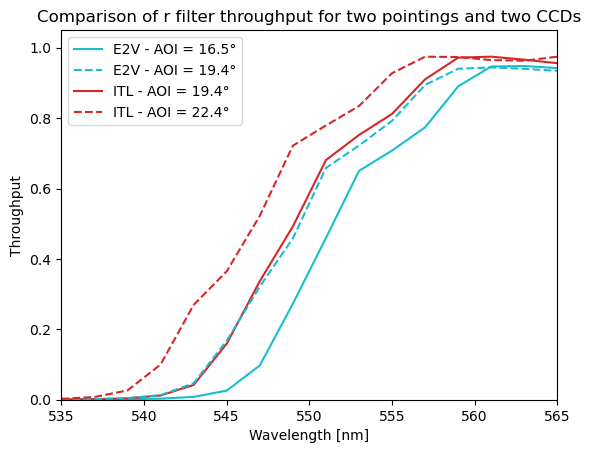}
    \includegraphics[width=0.52\linewidth]{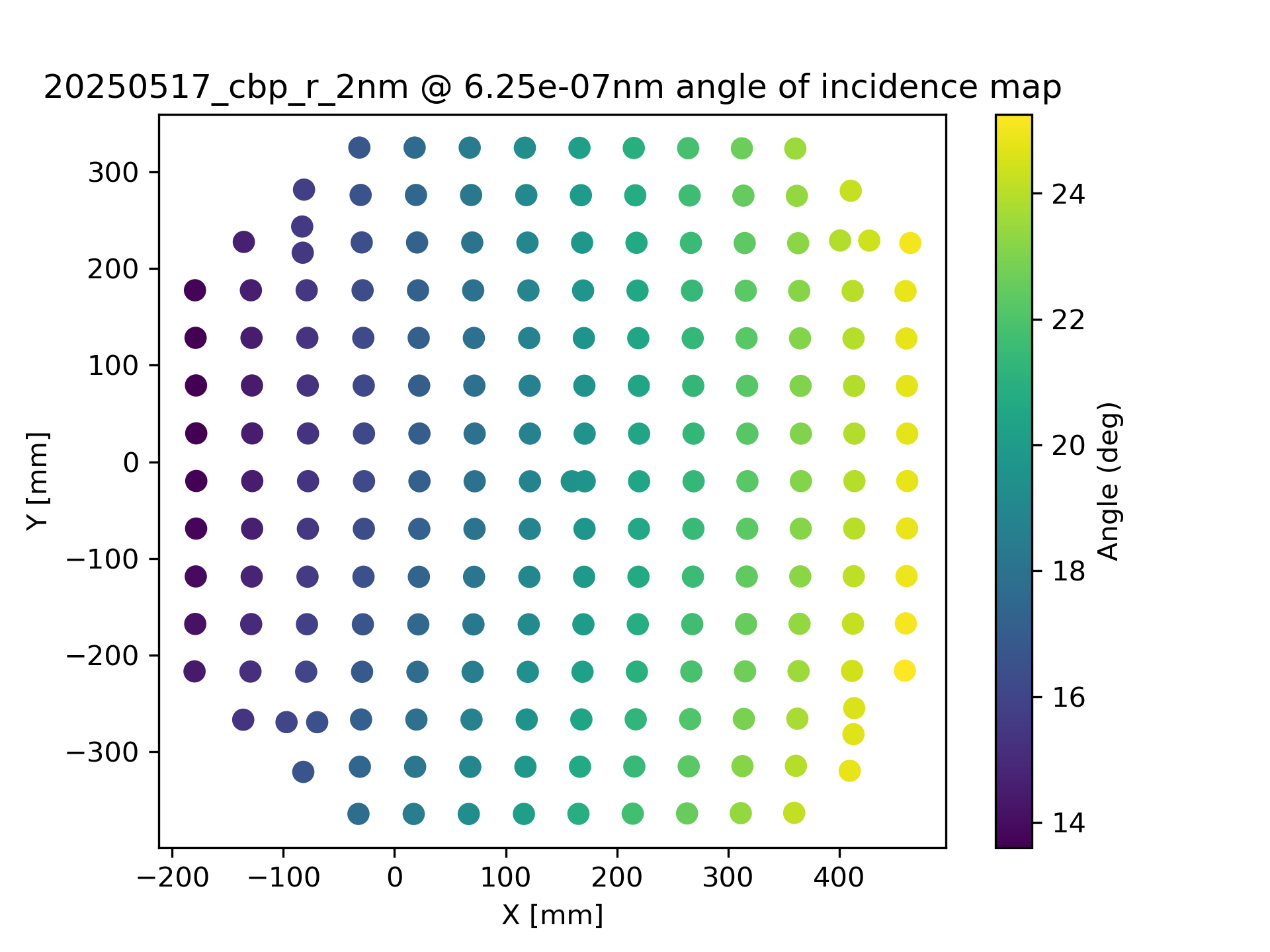}
    \caption{Left: Zoom on the blue edge of the $r$-band filter for one central CCD (e2v, blue) and one edge CCD (ITL, cyan), for two pointings (inner bottom and outer bottom of the mirror). AOI denotes for the angle of incidence values for each curve. Right: Simulated angle of incidence for each CBP spot on the filter plane as a function of focal-plane position, for a representative CBP pointing.}
    \label{fig:batoid_inc_angles}
\end{figure}

The LSST filters, manufactured by Materion, are 80-cm diameter curved interference filters consisting of multi-layer dielectric coatings. As a consequence of their curved geometry and multi-layer nature, the filter transmission varies as a function of both the spatial position and the angle of incidence of the incoming beam. The angle-of-incidence dependence manifests as a "blue-shift" of the bandpass edges toward shorter wavelengths as the incidence angle increases, as illustrated in the left panel of Figure~\ref{fig:batoid_inc_angles}. For on-sky observations with the Rubin Telescope, the angle of incidence on the filter is expected to range from approximately 14$^\circ$ to 23$^\circ$ between the inner and outer edges of the cone of illumination.

The angle of arrival of a CBP ray at the pupil (the primary mirror) determines where it lands on the Rubin focal plane. The radial location of the ray on the M1 pupil annulus determines the angle at which the ray passes through the filter. Since the CBP is located a considerable distance from the pupil, rays impinging  at different angles strike the pupil at different positions. This means that for a single CBP pointing, the arrival angle at the filter for rays that land at different locations on the focal plane are different. 
The resulting blue-shift can reach several nanometers to tens of nanometers for larger angle of incidence, so that the measured filter transmission edges exhibit spatial variation across the focal plane. Characterizing and modeling this effect is essential for spatially precise photometric calibration. Pupil-stitching that combines multiple CBP pointings can sample the full range of incidence angles for each focal plane location. 

An example of simulated ray incident angles on the filter plane for a representative CBP pointing, computed with the Batoid optical ray-tracing package \cite{batoid}, is shown in the right panel of Figure~\ref{fig:batoid_inc_angles}. The angle of incidence from the example CBP scan ranges from 14$^\circ$ to 25$^\circ$ which is slightly outside the on-sky range. 



\section{Conclusions}\label{section:conclusion}

We have presented early throughput results obtained with the Collimated Beam Projector at the Vera C. Rubin Observatory. The relative quantum efficiencies of the LSSTCam detectors have been measured with respect to the mean QE within each CCD family. A systematic percent-level discrepancy between the CBP-measured and tabulated relative RQE curves is observed across the 400–860\,nm range, highlighting the limitations of relying on laboratory data for per-mil photometric calibration. Achieving telescope plus camera response calibration at this level requires a more precise characterization of the CBP transmission and a vignetting model, which remains the primary outstanding ingredients.

Near-infrared fringing affects the detector response at wavelengths above 800\,nm at a level exceeding 5\%, introducing differential inter-detector variations that must be corrected or modeled to enable accurate photometric calibration in the $z$ and $y$ bands. This can be ameliorated by taking filter-in and filter-out data at the same wavelengths. 

Filter transmission profiles for five of the six LSST broadband filters have been measured from a single pupil pointing, providing the first in-situ characterization of the filters within the fully assembled Rubin system. An initial measurement of the angle-of-incidence-induced blue-shift of the filter edges has been performed. The next step for future analysis and full filter transmission measurement will be to establish a quantitative model of the blue-shifting effect with the LSST filter geometry.




\bibliography{report} 
\bibliographystyle{spiebib} 

\end{document}